\documentclass[onecolumn,usenatbib]{mn2e}
\usepackage{epsf}
\usepackage{epsfig}
\newcommand{\simgt}{\lower.5ex\hbox{$\; \buildrel > \over \sim \;$}}
\newcommand{\simlt}{\lower.5ex\hbox{$\; \buildrel < \over \sim \;$}}
\begin{document}
\title[The correlation of black hole mass 
with metallicity index]
{The correlation of black hole mass 
with metallicity index of host spheroid}
\author[S. Kisaka, Y.Kojima and Y. Otani]
{Shota Kisaka\thanks{E-mail:kisaka@theo.phys.sci.hiroshima-u.ac.jp}, 
Yasufumi Kojima\thanks{E-mail:kojima@theo.phys.sci.hiroshima-u.ac.jp}
and Yosuke Otani\\
Department of Physics, Hiroshima University,
Higashi-Hiroshima, 739-8526, Japan
}
\maketitle
\begin{abstract}
We investigate the correlation between
the mass of the supermassive black holes (SMBHs) and 
metal abundance, using existing data sets.
The SMBH mass  $M_{bh}$ is well correlated with 
integrated stellar feature of Mgb.
For 28 galaxies, the best-fit $M_{bh}$-Mgb relation 
has a small scatter, which is an equivalent level 
with other well-known relation, such as a correlation 
between the stellar velocity dispersion and the mass.
An averaged iron index $\langle $Fe$\rangle $ also 
positively correlates with $M_{bh}$, but
the  best-fit $M_{bh}$-$\langle $Fe$\rangle $ relation has 
a larger scatter. 
The difference comes from the synthesis and evolution mechanisms,
and may be important for the SMBH and star formation
history in the host spheroid. 
\end{abstract}
%
\section{INTRODUCTION}

  Recent observations of nearby massive spheroids 
(ellipticals, lenticular and spiral bulges) have established 
that supermassive black holes (SMBHs) are
present in the nuclei of galaxies. The mass $M_{bh}$ of SMBHs 
ranges from $10^{6}$ to $10^{9}M_{\odot }$. 
Several statistical correlations with characteristic parameters
of the host spheroids are explored: relation with
the luminosity $L$\citep{KR95,MH03,G07},
the mass $M_{s}$ of the spheroids\citep{Ma98,HR04}, 
the stellar velocity dispersion $\sigma $\citep{Ge00,FM00,Tr02},
the light concentration or S\'{e}rsic index $n$ \citep{Gr01,GD07},
gravitational binding energy $E_{g}$ \citep{AR07} and so on.
The recent updated survey for various galaxy parameters
is given in \citet{AR07}.
If the fitting models are good, the correlations may
provide some clues to the coevolution of SMBHs and the host spheroids.
Some tentative theories have been proposed to clarify the origin 
of these correlations, e.g., \citet{SR98,U01}.

 The galaxies have some aspects as an assemble of stars, 
gases and dark matters. 
The quantities $M_{s}$, $\sigma $, $E_{g}$ represent the 
dynamical aspect, while $L$, $n$ represent the photometric
one, although they are closely related each other.
Very little attention is paid to the chemical one. 
The observation is not so easy and therefore the correlation between
SMBH mass and spheroid chemical property is not well studied so far.
Both the metal abundance and black hole mass increase with time 
and we therefore expect some correlations between them. 
It is important to clarify the
correlation, and the extent, if any. The chemical parameter 
using Lick/IDS absorption line indices, e.g., \citet{Wo94}, 
especially Mg and Fe is systematically studied 
in literatures\citep{Tr00,De05,Ku06} for the galaxies
in which presence of SMBH is reliable. Using the published data,
we investigate the correlation with SMBH mass in this paper.

  As a sample we select only 28 galaxies, for which SMBH masses
are measured by dynamical methods. 
In section 2, we describe the sample of galaxies and
criterion of our choice.  In section 3, we derive the 
correlation between
$M_{bh}$ and $\sigma $, and that between $M_{bh}$ and 
the B-band magnitude $M_{B}$ from our sample. 
These correlations are known to be good, and are
used in order to check no serious bias involved in the sample. 
Our results for $M_{bh}$-$\sigma $ and $M_{bh}$-$M_{B}$ relations 
are not so different from those by \citet{AR07,G07}, although the source sample
is not the same.  We also show the correlation between black hole mass
and metal abundance index using our sample.
Finally, a discussion of our results is given in section 4.

\section{THE SAMPLE}
{\footnotesize 
\begin{tabular}{ccccccccc}
\multicolumn{9}{c}{TABLE 1 Galaxy Parameter Used in The Fits} \\ \hline
Galaxy & Type$^{a}$ & $M_{bh}$ (low,high) & Reference & $\sigma ^{b}$ & $%
M_{B}$ $^{c}$ & Mgb & $\langle Fe\rangle $ & Reference \\ 
&  & ($M_{\odot }$) &  & (km s$^{-1}$) & (mag) & (mag) & (mag) &  \\ 
(1) & (2) & (3) & (4) & (5) & (6) & (7) & (8) & (9) \\ \hline
NGC 221 & S0 & 2.5E6 (2.0,3.0) & 1 & 75 & -15.3 & $0.099\pm 0.001$ & $%
0.078\pm 0.001$ & 19,20 \\ 
NGC 224 & Sb & 1.4E8 (1.1,2.3) & 2 & 160 & -19.0 & $0.175\pm 0.002$ & $%
0.087\pm 0.001$ & 20 \\ 
NGC 821 & E4 & 8.5E7 (5.0,12) & 3 & 209 & -20.4 & $0.159\pm 0.002$ & $%
0.086\pm 0.001$ & 19,20,21 \\ 
NGC 1023 & SB0 & 4.4E7 (3.9,4.9) & 4 & 205 & -18.4 & $0.162\pm 0.003$ &%
$\cdots$& 21 \\  
NGC 2974 & E4 & 1.7E8 (1.5,1.9) & 5,6 & 233 & -20.9 & $0.159\pm 0.002$ & $%
0.084\pm 0.002$ & 19,21 \\ 
NGC 3115 & S0 & 1.0E9 (0.4,2.0) & 7,8 & 230 & -20.2 & $0.177\pm 0.002$ & $%
0.096\pm 0.001$ & 19,22 \\ 
NGC 3245 & S0 & 2.1E8 (1.6,2.6) & 9 & 205 & -19.6 & $0.161\pm 0.006$ & $%
0.077\pm 0.002$ & 19 \\ 
NGC 3377 & E5 & 1.0E8 (0.9,1.9) & 10 & 145 & -19.0 & $0.142\pm 0.002$ & $%
0.076\pm 0.001$ & 19,20,21 \\ 
NGC 3379 & E1 & 1.4E8 (0.6,2.0) & 11 & 206 & -19.9 & $0.168\pm 0.001$ & $%
0.084\pm 0.001$ & 19,20,21 \\ 
NGC 3384 & SB0 & 1.6E7 (1.4,1.7) & 10 & 143 & -19.0 & $0.154\pm 0.002$ & $%
0.075\pm 0.001$ & 19,21 \\ 
NGC 3414 & S0 & 2.5E8 (2.2,2.8) & 5,6 & 205 & -20.0 & $0.170\pm 0.003$ & $%
0.081\pm 0.001$ & 19,21 \\ 
NGC 3608 & E2 & 1.9E8 (1.3,2.9) & 10 & 182 & -19.9 & $0.162\pm 0.002$ & $%
0.085\pm 0.001$ & 19,20,21 \\ 
NGC 4261 & E2 & 5.2E8 (4.1,6.2) & 12 & 315 & -21.1 & $0.186\pm 0.002$ & $%
0.088\pm 0.001$ & 19,20 \\  
NGC 4374 & E1 & 4.6E8 (2.8,8.1) & 13,14 & 296 & -21.4 & $0.168\pm 0.001$ & $%
0.082\pm 0.002$ & 19,21,22 \\ 
NGC 4459 & S0 & 7.0E7 (5.7,8.3) & 15 & 186 & -19.1 & $0.139\pm 0.003$ & $%
\cdots$ & 21 \\ 
NGC 4473 & E5 & 1.1E8 (0.3,1.5) & 10 & 190 & -19.9 & $0.168\pm 0.002$ & $%
0.088\pm 0.002$ & 21,22 \\ 
NGC 4486 & E0 & 3.0E9 (2.0,4.0) & 7,16,17 & 375 & -21.5 & $0.199\pm 0.002$ & 
$0.086\pm 0.002$ & 21,22 \\ 
NGC 4552 & E & 5.0E8 (4.5,5.5) & 5,6 & 252 & -19.2 & $0.185\pm 0.002$ & $%
0.084\pm 0.001$ & 20,21 \\ 
NGC 4564 & S0 & 5.6E7 (4.8,5.9) & 10 & 162 & -17.4 & $0.173\pm 0.003$ & $%
\cdots$ & 21 \\ 
NGC 4596 & SB0 & 7.8E7 (4.5,11.6) & 15 & 152 & -20.6 & $0.150\pm 0.008$ & $%
\cdots $ & 23 \\ 
NGC 4621 & E5 & 4.0E8 (3.6,4.4) & 5,6 & 211 & -19.5 & $0.178\pm 0.002$ & $%
0.088\pm 0.002$ & 21,22 \\ 
NGC 4649 & E1 & 2.0E9 (1.4,2.4) & 10 & 385 & -21.3 & $0.194\pm 0.002$ & $%
0.085\pm 0.001$ & 20 \\ 
NGC 4697 & E4 & 1.7E8 (1.6,1.9) & 10 & 177 & -20.2 & $0.146\pm 0.002$ & $%
0.078\pm 0.001$ & 20 \\ 
NGC 5813 & E1 & 7.0E8 (6.3,7.7) & 5,6 & 230 & -20.9 & $0.167\pm 0.001$ & $%
0.078\pm 0.001$ & 19,20,21 \\ 
NGC 5845 & E3 & 2.4E8 (1.0,2.8) & 10 & 234 & -18.7 & $0.169\pm 0.003$ & $%
0.093\pm 0.002$ & 19,21 \\ 
NGC 5846 & E0 & 1.1E9 (1.0,1.2) & 5,6 & 238 & -21.2 & $0.179\pm 0.001$ & $%
0.085\pm 0.001$ & 19,20,21,22 \\ 
NGC 7052 & E4 & 3.3E8 (2.0,5.6) & 7,18 & 266 & -21.2 & $0.182\pm 0.002$ & $%
0.080\pm 0.002$ & 20 \\  
NGC 7457 & S0 & 3.5E6 (2.1,4.6) & 10 & 67 & -17.0 & $0.101\pm 0.003$ & $%
\cdots$ & 21 \\ \hline
\multicolumn{9}{l}{}%
\end{tabular}

NOTES.-Col.(1):Galaxy name. Col.(2):Galaxy type. Col.(3):SMBH mass.
Col.(4):Reference for col.(3). Col.(5):Effective stellar bulge velocity
dispersion. Col.(6):Absolute bulge B-band magnitude. Col.(7),(8)Line
strength index measurements of the luminosity weighted spectrum within $%
r_{e}/8$ for the Mgb and $\langle$Fe$\rangle$ indices. Col.(9):Reference for
col.(7),(8).

$^{a}$Morphological type from NED except for NGC 224 and NGC
4564. The type given here for those galaxies is from \cite{G02} for NGC
221, and from \cite{Tr04}
for NGC 4564.

$^{b}$The value is taken from \cite{Tr02} except for NGC 2974, NGC 3414,
NGC 4552, NGC 4621, NGC 5813, NGC5846 \cite{H08}, NGC 4374 \cite{MH03}.

$^{c}$The value is taken from \cite{G07} or extracted from \cite{MH03}
and the Third Reference Catalogue of Bright Galaxies \cite{de91}.

REFERENCES.-(1)\cite{Ve02} (2)\cite{Be05} (3)\cite{Ri04} (4)\cite{Bo01}
(5)\cite{Ca07} (6)\cite{H08} (7)\cite{Tr02} (8)\cite{Ko96} (9)\cite{Ba01}
(10)\cite{Ge03} (11)\cite{Sh06} (12)\cite{Fe96}
(13)\cite{MB01} (14)\cite{KG01} (15)\cite{Sa01} (16)\cite{Ha94} (17)\cite{Ma97}
(18)\cite{vv98} (19)\cite{De05} (20)\cite{Tr00}
(21)\cite{Ku06} (22)\cite{H05} (23)\cite{Pe07}
}
\bigskip\bigskip

%
  Our sample consists of 28 nearby galaxies, which are listed 
in Table 1. We put two criteria to pick up the sample 
from literatures. 
First, the BH masses are measured by good spatial resolution.
The sources are limited to
$M_{bh} > \sigma^{2} r_{res}/(2G) $, 
where $\sigma^{2} $ is stellar velocity dispersion,
$r_{res}$ the instrumental spatial resolution. 
This condition is used by e.g., \cite{MH03,H08}.
Second, metal indices of Mg or Fe are measured. 
Most of our sample corresponds to that of \cite{Tr02}.  
21 galaxies are drawn from their table
and 7 galaxies are added:
NGC 2974, NGC 3414, NGC 4552, NGC4621, NGC 5813 and NGC5846
from \cite{Ca07,H08},
and NGC 4374 from \cite{MB01}. 
The masses of SMBHs are updated for 3 galaxies,
NGC 224\citep{Be05}, NGC 821\citep{Ri04} and NGC 3379\citep{Sh06}. 
The galaxy morphology class is also listed in Table 1
from the NASA Extragalactic Database (NED).
However, some corrections are added to the data: 
NGC 4564\citep{Tr04} and NGC 221\citep{G02}  are 
now recognized as S0 galaxies.
In Table 1, effective velocity dispersion $\sigma $ is tabulated 
as an indicator of the dynamical parameter of spheroids. 
Following \cite{Tr02}, the relative errors in $\sigma $ are 
estimated as 5 \% (0.021 dex). As the
photometric parameter of spheroids, the B-band magnitude $M_{B}$
is tabulated.
Following \cite{G07}, the relative errors in $M_{B}$ are 0.3 mag. 
For the chemical parameter of spheroids, we use spectral absorption 
line indices in the visual as defined 
by the Lick group, e.g., \cite{Bu84,Fa85}. 
Lick indices have proven to be a useful tool for the derivation
of ages and metallicity of unresolved stellar populations\citep{De05}.
In this paper, we use Mgb, Fe5270, Fe5335 indices, measuring
respectively the strength of MgI at
$\lambda \simeq 5156-5197$\AA, 
FeI at $\lambda \simeq 5246-5286$\AA  and FeI 
at $\lambda \simeq 5312-5352$\AA. 
These indices in mag are calculated
by the standard equations: 
\begin{equation}
EW_{mag}=-2.5\log \biggl\{\frac{\int_{\lambda _{1}}^{\lambda _{2}}[F(\lambda
)/C(\lambda )]d\lambda }{\lambda _{2}-\lambda _{1}}\biggr\},
\end{equation}%
where 
\begin{equation}
C(\lambda )=F_{b}\frac{\lambda _{r}-\lambda }{\lambda _{r}-\lambda _{b}}%
+F_{r}\frac{\lambda -\lambda _{b}}{\lambda _{r}-\lambda _{b}}
\end{equation}%
and $\lambda _{b}$ and $\lambda _{r}$ are the mean wavelength in the blue
and red pseudo-continuum intervals, respectively. We have adopted the
spectral pseudocontinua and band passes of the Mgb, Fe5270, Fe5335 Lick/IDS
indices defined in \cite{Wo94}. We use a combined
\textquotedblleft iron\textquotedblright\ index, $\langle $Fe$\rangle $
defined by 
\begin{equation}
\langle Fe\rangle \equiv \frac{1}{2} (Fe5270+Fe5335).
\end{equation}
 This index has smaller random error than either that of Fe5270 or that
of Fe5335 \citep{Go90}. 
These index values used from the literatures,
are the measurements within the central region
 $r < r_{e}/8$, where $r_{e}$ is effective radius of the spheroids. 
The properties of the spheroids and SMBHs are 
listed with their references in Table 1.

\section{RESULTS}

In order to fit the data to the linear relation $y=a+bx$, 
we use a version of the routine FITEXY \citep{Pr92} 
modified by \cite{Tr02}.
\cite{No06} showed that this algorithm is the most efficient
and least biased among a set of algorithms explored by them.
The best-fit values $a$ and $b$ are calculated from data 
$(x_{i},y_{i}) (i=1,\cdots,N)$ by minimizing the quantity 
\begin{equation}
\chi ^{2}=\sum_{i=1}^{N}\frac{(y_{i}-a-bx_{i})^{2}}{\delta
y_{i}^{2}+\epsilon ^{2}+b^{2}\delta x_{i}^{2}},
\end{equation}%
where $\delta x_{i}$ and $\delta y_{i}$ are the measurement 
uncertainty in $x_{i}$ and $y_{i}$.  
We use a constant value of 5 \% (0.021 dex) 
for velocity dispersion $\sigma $, 
and 0.3 mag for the B-band magnitude $M_{B}$.
The uncertainty for the metal abundance is listed in Table 1, 
and is used for each source.
The intrinsic scatter $\epsilon $ in the linear relation
is calculated when the reduced chi-squared value 
$\chi ^{2}/(N-2)$ is equal to 1, 
after many trails to fit it for various $\epsilon $.
The upper and lower uncertainties of $\epsilon $ are estimated 
by the values for $\chi ^{2}/(N-2) = 1\pm \sqrt{2/N}$. 

\subsection{Dynamical and Photometric Properties}

In order to check the dependence on the sampling data, we 
derive $M_{bh}$-$\sigma $ and $M_{bh}$-$M_{B}$ relations
from our sample. 
First, for the dynamical aspect, we show a relation between
the SMBH mass and the stellar velocity dispersion derived from
Table 1.
Our result of $M_{bh}$-$\sigma$ relation is given by
\begin{equation}
\log (M_{bh}/M_{\odot })=(3.98\pm 0.35)\log (\sigma/200)+8.23\pm 0.06
\label{eqnbhsigma1}
\end{equation}
with $\epsilon=0.25_{-0.03}^{+0.06}$dex in $\log M_{bh}$.
This relation is consistent with previous relations.
For example, recent updated one by \cite{AR07} 
using a sample of 23 galaxies is 
\begin{equation}
\log (M_{bh}/M_{\odot })=(3.79\pm 0.32)\log (\sigma/200)+8.16\pm 0.06
\label{eqnbhsigma2}
\end{equation}
with $\epsilon=0.23$ dex.
Figure 1 shows our $M_{bh}$-$\sigma$ relation (\ref{eqnbhsigma1})
in comparison with (\ref{eqnbhsigma2}).

Next, for the photometric aspect, we
derive the $M_{bh}$-$M_{B}$ relation. Our result is 
\begin{equation}
\log (M_{bh}/M_{\odot })=(-0.42\pm 0.05)(M_{B}+19.5)+8.14\pm 0.08
\label{eqnbhb1}
\end{equation}
with $\epsilon=0.36_{-0.05}^{+0.07}$dex\footnote{ 
We have also derived the relation of $M_{bh}$ with the
K-band magnitude, but there is no significant 
difference in the scatters of both relations.
}
\cite{G07} obtained 
\begin{equation}
\log (M_{bh}/M_{\odot })=(-0.40\pm 0.05)(M_{B}+19.5)+8.27\pm 0.08
\label{eqnbhb2}
\end{equation}
with  $\epsilon=0.30_{-0.05}^{+0.04}$dex.
Although intrinsic scatter in our relation is a little high, 
both two relations agree within the uncertainties.
Both  relations (\ref{eqnbhb1}) and (\ref{eqnbhb2}) 
are plotted in Figure 2.
By comparing with previous results, 
we may say that there is no serious bias in our sampling data.

\begin{figure}
 \begin{minipage}{0.45\linewidth}
  \begin{center}
   \includegraphics[width=50mm,angle=270]{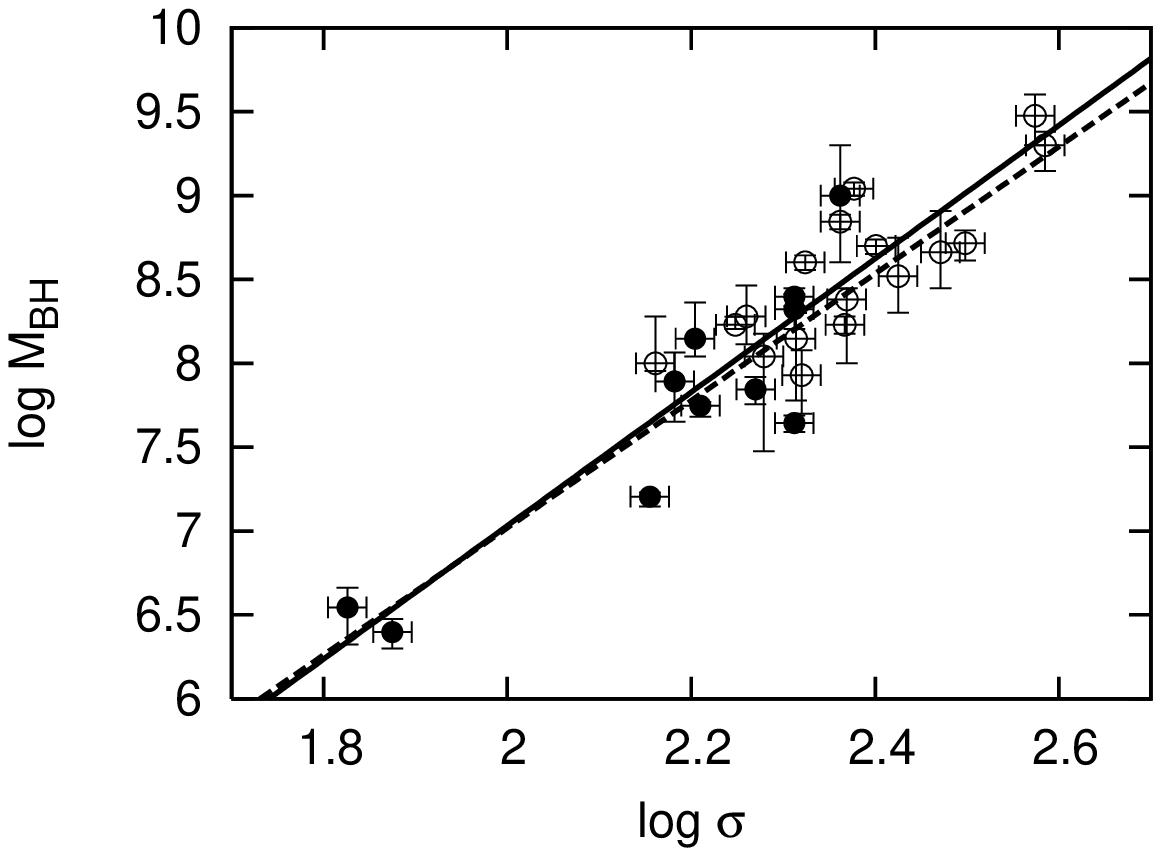}
   \caption{
 Relationship between SMBH mass and effective stellar
 velocity dispersion for 28 galaxy samples. 
 The solid line represents our best fit relation 
 (\ref{eqnbhsigma1}).
 For a comparison, the relation by \citep{AR07}
 is shown by  a  dashed line. 
 Elliptical galaxies are denoted by open circles, 
 lenticulars and spirals by filled circles.
}
  \end{center}
 \end{minipage}
\hspace{0.05\linewidth}
 \begin{minipage}{0.45\linewidth}
  \begin{center}
   \includegraphics[width=50mm,angle=270]{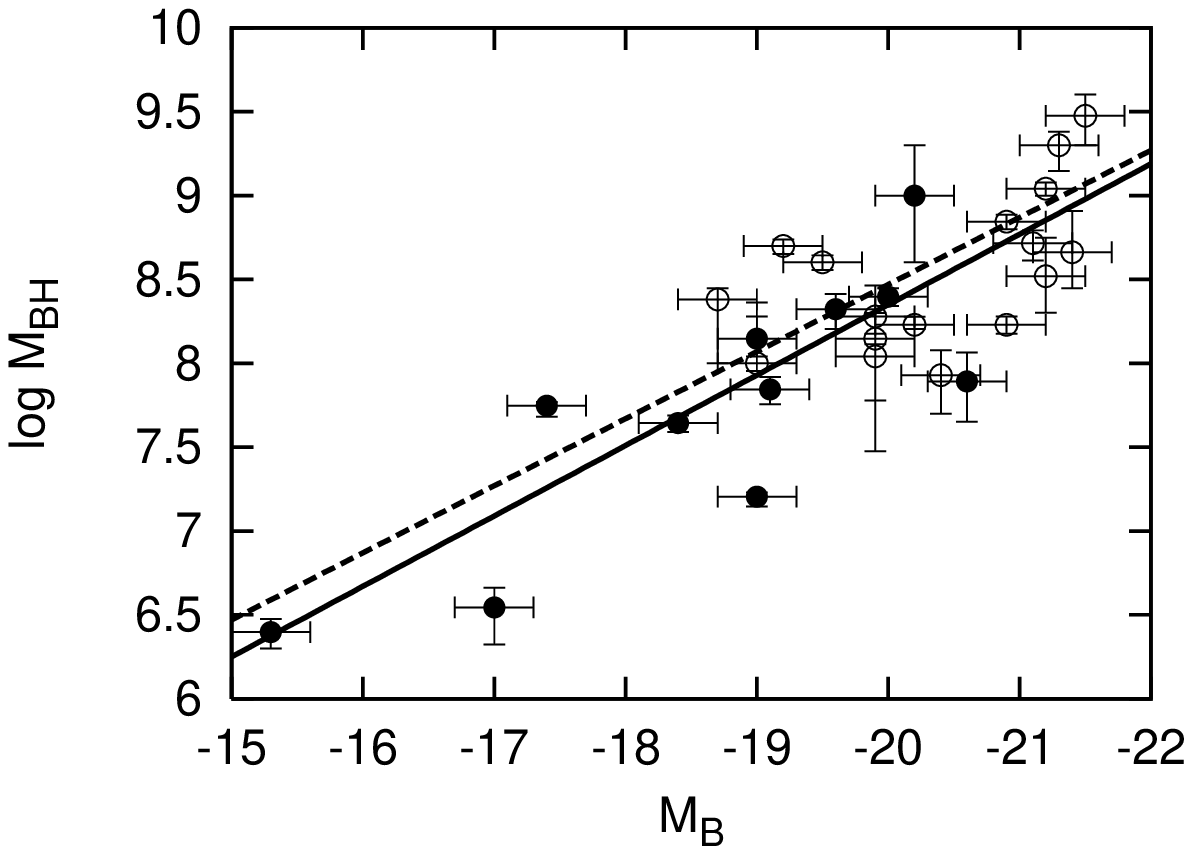}
\caption{
 Relationship between SMBH mass and  absolute B-band luminosity
 of the spheroid  for 28 galaxy samples. 
 The solid line represents our best fit relation (\ref{eqnbhb1}). 
 For a comparison, the relation found by \citep{G07}
 is shown by a dashed line. The symbols are the same as Figure 1.
}
  \end{center}
 \end{minipage}
\end{figure}

\subsection{Chemical Property}

In Figure 3, we plot $M_{bh}$ as a function of
the integrated stellar feature of Mgb for the 28 galaxies.
It is found that the mass $M_{bh}$ significantly
correlates with the index Mgb. 
The best fit relation is obtained as 
\begin{equation}
\log (M_{bh}/M_{\odot })=(27.23\pm 3.00)(Mgb-0.16)+8.13\pm 0.07
\end{equation}
with $\epsilon=0.32_{-0.04}^{+0.06}$dex.
The scatter is not so bad as that of well-fitted
$M_{bh}$-$\sigma$ and $M_{bh}$-$M_{B}$ relations.
The index Mgb is a good indicator, but
other indicator of the metal abundance is not so good.
We have tested the correlation with different index. 
For instance, we show $M_{bh}$ as a function of
$\langle$Fe$ \rangle $ index in Figure 4.
The best fit relation is 
\begin{equation}
\log (M_{bh}/M_{\odot })=(60.25\pm 26.01)(\langle Fe\rangle
 -0.083)+8.34\pm 0.13
\end{equation}
with $\epsilon=0.58_{-0.07}^{+0.12}$ dex.
This relation has much larger scatter than that of eqs.(5)-(9).
It is clear from Figure 4 that the index
$\langle$Fe$ \rangle $ is not good one.
Summarizing the relationships obtained in our own study, 
we find that 
$M_{bh}$-$\sigma$ is the best,
$M_{bh}$-Mgb and $M_{bh}$-$M_{B}$ are moderate,
and $M_{bh}$-$\langle$Fe$ \rangle $ is the worst.
\begin{figure}
 \begin{minipage}{0.45\linewidth}
  \begin{center}
   \includegraphics[width=50mm,angle=270]{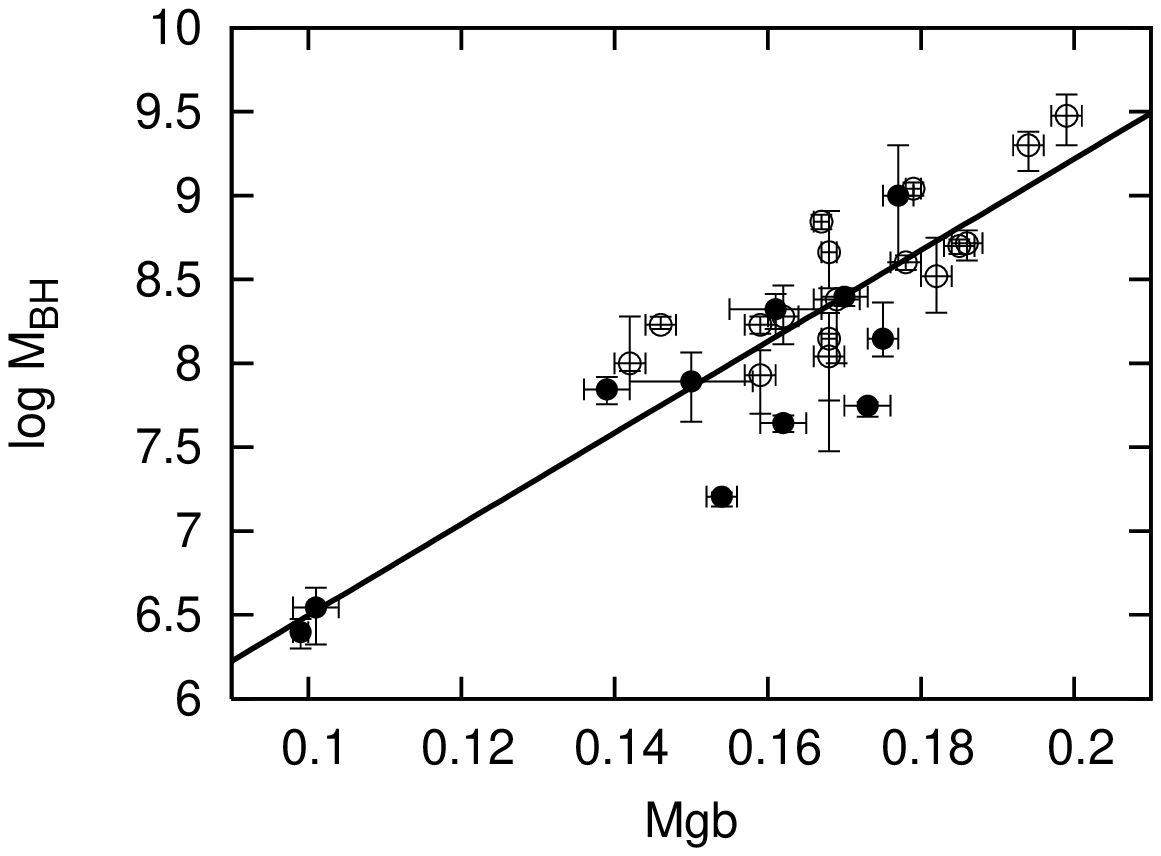}
\caption{
 Relationship between SMBH mass and Mgb index value within a circular
 aperture of $r_{e}/8$ for our samples. The symbols and line are 
 the same as in Figure 1.}
  \end{center}
 \end{minipage}
%
\hspace{0.01\linewidth}
%
 \begin{minipage}{0.45\linewidth}
  \begin{center}
   \includegraphics[width=50mm,angle=270]{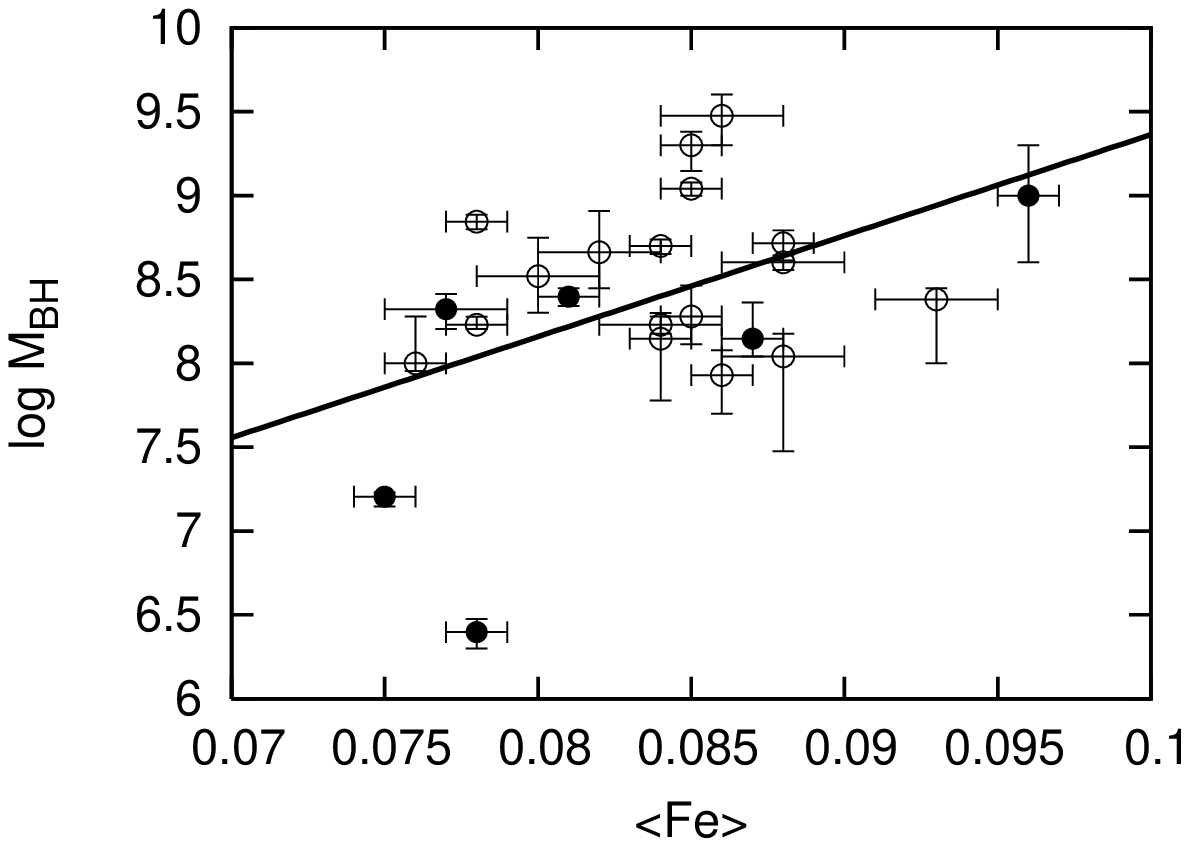}
\caption{
 Relationship between SMBH mass and $\langle$Fe$\rangle$ index
 within $r_{e}/8$ for our samples. The symbols and line are
 the same  as in Figure 1.}
  \end{center}
 \end{minipage}
\end{figure}

\section{DISCUSSION}

  We have shown that a good correlation between 
SMBH mass and Mgb index value.
The best-fitting $M_{bh}$-Mgb relation has small
intrinsic scatter 0.32 dex which is comparable one in other 
strong correlations found so far.  
Such a new correlation in the chemical aspect
is expected through other relations in dynamical/photometric
aspects. The metal abundance is roughly correlated with
total stellar mass, absolute magnitudes etc. e.g., \cite{Pa97}.
In particular, there is a remarkably tight relation between
Mg$_2$ and the central velocity dispersion of stars\citep{Te81,Be93}.
The positive correlations have been found 
between some quantities characterizing the hosts
and SMBH mass.
A positive correlation is also suggested between the SMBH mass
and the metallicity derived from emission line ratios in 578 AGNs
spanning a wide range in redshifts\citep{Wa03}. 
Our result of Mgb for nearby galaxies is more tight.
Thus, positive correlation is expected, but the degree was not 
clear beforehand. It was not clear which indicators of the 
metal abundance strongly correlate with the SMBH mass.
The index $\langle$Fe$\rangle$ correlates with it,
but the intrinsic scatter is not so small as that of Mgb.
Heavy elements Mg and Fe are synthesized by two different types of 
supernovae (Type Ia and II), with different time scales. 
This evolutionary difference causes the scatter
in the correlations.

  The growth of black hole mass is mainly determined by the
accretion rate and the lifetime of the activity.
The environmental factors near the central region of galaxies 
may partially be affected by some global quantities,
such as the mass and size of the host.
If the mass $M_{bh}$ of SMBH is determined solely by
the spheroid mass $M_s$  
as $M_{bh} = \varepsilon  M_s$ $(\varepsilon \sim 10^{-5})$,
then we have  
$M_{bh} \propto M_s$ $ \propto L$ $\propto  \sigma ^4$,
where we assume that  a constant mass-to-light ratio and the 
Faber-Jackson relation in elliptical galaxies hold in the spheroids.
The features in the host spheroids are
transferred to the relations with the black hole mass,
$M_{bh}$-$L$ and $M_{bh}$-$\sigma$ relations.
Other features in the hosts, binding energy, light concentration
and so on also give some relations with $M_{bh}$.
\cite{Be93} discuss that
the strength of Mg$_2$ is determined not only by
the global mass $M_s$, but also by the local stellar density,
which is related with the star-formation rate etc.
It is therefore important to examine 
the correlation of $M_{bh} $ with the other physical quantities
of the spheroid, in addition to $M_s$. 
The metal abundance is a tracer for integrated stellar populations.
The tight relation $M_{bh}$-Mgb, which is discovered here
but is still tentative,  
may be useful for the better understanding of the coevolution
of SMBH and host spheroid, if it is not accidental.
Further work is needed to clarify 
whether this relation is fundamental or not.
%

\section*{Acknowledgments}
 This research has made use of the NASA/IPAC Extragalactic Database (NED),
which is operated by the Jet Propulsion Laboratory, California Institute of
Technology, under contract with the National Aeronautics and Space
Administration.
This work was supported in part by the Grant-in-Aid for
Scientific Research (No.16540256) from
the Japanese Ministry of Education, Culture, Sports,
Science and Technology.


\end{document}